\documentclass[a4paper]{article}
\usepackage{onecolceurws}
\usepackage{mathptmx}
\usepackage[T1]{fontenc}
\usepackage[utf8]{inputenc}
\usepackage{ifthen}
\usepackage[english]{babel}
\usepackage{graphicx}
\usepackage{color}
\usepackage{isabelle,isabellesym}
\usepackage{pdfsetup}

\isabellestyle{literal}

\isadroptag{theory}

\definecolor{linkcolor}{rgb}{0,0,0.5}
\hypersetup{colorlinks=true,linkcolor=linkcolor,citecolor=linkcolor,filecolor=linkcolor,urlcolor=linkcolor,pdfpagelabels}

\renewcommand{\isastyletxt}{\isastyletext}

\newcommand{\secref}[1]{\S\ref{#1}}
\newcommand{\figref}[1]{fig.~\ref{#1}}

\makeatletter
\renewcommand\paragraph{\@startsection{paragraph}{4}{\z@}%
{1.25ex \@plus1ex \@minus.2ex}%
{-1ex}%
{\normalfont\normalsize\bfseries}}
\makeatother

\begin{document}

\title{Isabelle technology for the Archive of Formal Proofs \\
  with application to MMT}
\author{Makarius Wenzel}
\institution{\url{https://sketis.net}}
\maketitle

\begin{abstract}

This is an overview of the Isabelle technology behind the Archive of Formal
Proofs (AFP). Interactive development and quasi-interactive build jobs impose
significant demands of scalability on the logic (usually Isabelle/HOL), on
Isabelle/ML for mathematical tool implementation, and on Isabelle/Scala for
physical system integration --- all integrated in Isabelle/PIDE (the Prover
IDE). Continuous growth of AFP has demanded continuous improvements of
Isabelle performance. This is a report on the situation in Isabelle2019 (June
2019), with notable add-ons like prover session exports and headless PIDE for
automated updates based on semantic information. An example application is
Isabelle/MMT, which is able to turn all of Isabelle + AFP into OMDoc and RDF
triples, but it is straight-forward to reuse the Isabelle technology for other
applications.

\end{abstract}

\begin{isabellebody}%
\setisabellecontext{Paper}%
\isadelimtheory
\isanewline
\isanewline
\endisadelimtheory
\isatagtheory
\isacommand{theory}\isamarkupfalse%
\ Paper\isanewline
\ \ \isakeyword{imports}\ Main\isanewline
\isakeyword{begin}%
\endisatagtheory
{\isafoldtheory}%
\isadelimtheory
\endisadelimtheory
\isadelimdocument
\endisadelimdocument
\isatagdocument
\isamarkupsection{Motivation: scalability for Isabelle/AFP \label{sec:motivation}%
}
\isamarkuptrue%
\endisatagdocument
{\isafolddocument}%
\isadelimdocument
\endisadelimdocument
\begin{isamarkuptext}%
The Archive of Formal Proofs\footnote{\url{https://www.isa-afp.org}} is a repository
of formalized mathematics that is organized like a scientific journal. The
maintenance model of Isabelle + AFP is conservative in the sense that
changes to the Isabelle system and basic Isabelle/HOL libraries are always
``pushed through'' to applications in AFP: this demands reasonably fast
feedback from build jobs. The following practical \emph{time scales} for testing
have emerged as a rule of thumb:

\begin{description}%
\item [Online time (max. 45min)] for quasi-interactive builds while sitting at
the computer and doing other things. This time span is also anecdotal as
the \emph{Paris commuter's constant}, i.e. the practical limit of a person
sitting patiently on a train to wait for its arrival.

\item [Offline time (max. 2h)] for batch-builds while being absent and not
watching it. This is the time span of a classic \emph{French lunch break}.%
\end{description}

\noindent Successful trimming of build times to these limits makes maintenance and
development of new (and larger) AFP articles feasible. Thus its growth can
continue unhindered, but the demand for performance increases! Ultimately,
we cannot win this race of the technology versus cumulative applications,
but we can see how large and prosperous our mathematical library can get,
before eventual stagnation.

\medskip Current AFP\footnote{Repository \url{https://bitbucket.org/isa-afp/afp-2019} version
  841f0dcedae1 from 08-Jun-2019.} has 320 authors, 473 articles, 4955
theories, $10^5$ theorem statements, $10^6$ internal facts, $10^8$ bytes of
text. The sustained growth of AFP (since its foundation in 2004) is
illustrated in \figref{fig:afp-statistics}: the diagram shows the source
text size with the date of the first appearance of an article (existing
articles are sometimes extended later). The build time on a high-end server
with many cores and fast memory is as follows (using 8 processes with 8
threads each); here the special group of \isatt{very{\char`\_}slow} sessions is always
excluded, and further session groups are selected as given below:

\begin{itemize}%
\item Isabelle with \isatt{main} sessions only: 7.5min elapsed time, 53min CPU time
(factor 7.0)

\item Isabelle with all sessions: 12min elapsed time, 5h04 CPU time (factor
25.0)

\item AFP without \isatt{slow} / \isatt{large} sessions: 51min elapsed time, 25h47 CPU
time (factor 30.3)

\item AFP with \isatt{slow} / \isatt{large} sessions only: 50min elapsed time, 12h04 CPU
time (factor 14.5)

\item Isabelle + AFP with all sessions: 1h14 elapsed time, 42h11 CPU time
(factor 34.2)%
\end{itemize}

Typical maintenance proceeds as follows. Changes of the main Isabelle
libraries are quickly tried out on their own: 7.5min or 12min. Then follows
the regular AFP test without \isatt{slow} / \isatt{large} sessions: 51min is slightly
above the ``Paris commuter's constant'', but still bearable. The \isatt{slow} /
\isatt{large} sessions themselves are at the same order, but traditionally
postponed to a nightly build. Alternatively, a simultaneous test of Isabelle
+ AFP runs 1h14 on this high-end hardware: on lesser machines it is closer
to the ``French lunch break'' of 2h.

Overall, the current situation is within the expected parameters: it means
that the present Isabelle technology will allow AFP to prosper and grow a
bit further, but the technology needs to follow up eventually.

\begin{figure}[h!tb]
\begin{center}
\includegraphics[width=0.95\textwidth]{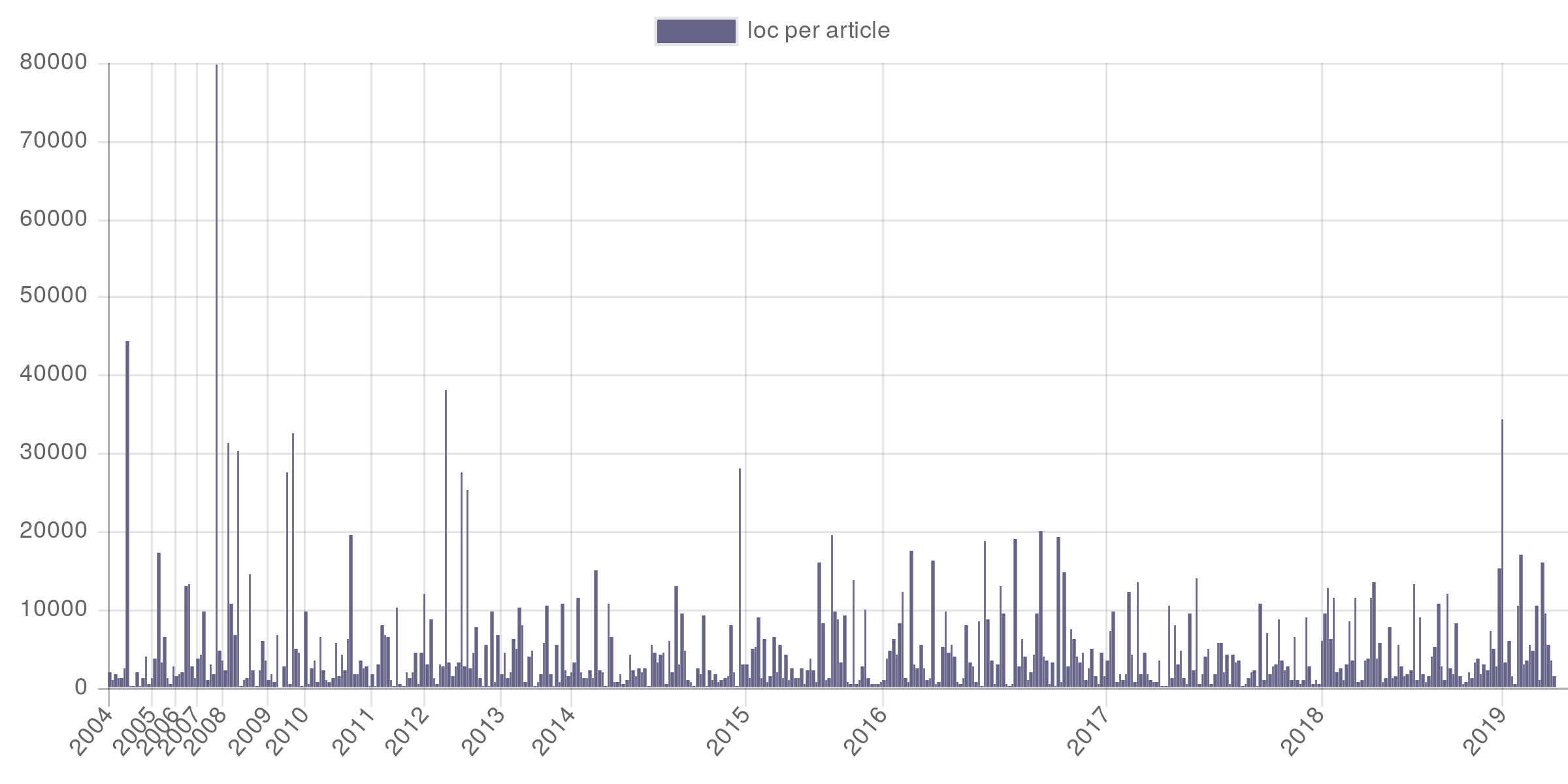}
\end{center}
\caption{Text size of AFP articles wrt. date of first appearance}
\label{fig:afp-statistics}
\end{figure}%
\end{isamarkuptext}\isamarkuptrue%
\isadelimdocument
\endisadelimdocument
\isatagdocument
\isamarkupsection{Isabelle technology%
}
\isamarkuptrue%
\endisatagdocument
{\isafolddocument}%
\isadelimdocument
\endisadelimdocument
\begin{isamarkuptext}%
Isabelle\footnote{\url{https://isabelle.in.tum.de}} was historically introduced as
  ``logical framework'' or ``generic proof assistant'' \cite{paulson700}, but
  over the decades it has evolved into software technology to support large
  libraries of formal mathematics.

  Usually the object-logic is Isabelle/HOL, because that has the
  best-developed collection of theories and tools (e.g. derived specifications
  and proof methods), but that is somewhat accidental. Isabelle as a platform
  remains open to other logics: it is merely a matter to cultivate sufficient
  library support for applications (which can take many years). Subsequently
  we shall ignore logic and talk about the Isabelle infrastructure. The
  Isabelle platform is organized as distinctive sub-systems, with a canonical
  naming scheme of ``Isabelle/XYZ'', notably Isabelle/ML as language for
  \emph{internal tool implementation}, Isabelle/Scala as language for \emph{external
  system integration}, and Isabelle/PIDE as framework for \emph{semantic
  interaction} (with human users or other tools).%
\end{isamarkuptext}\isamarkuptrue%
\isadelimdocument
\endisadelimdocument
\isatagdocument
\isamarkupparagraph{Isabelle/ML%
}
\isamarkuptrue%
\endisatagdocument
{\isafolddocument}%
\isadelimdocument
\endisadelimdocument
\begin{isamarkuptext}%
is a distinctive member of the ML family, with a rich library, source-level
debugger, and high-end IDE. Since April 2016, the IDE can load the
Isabelle/ML/Pure bootstrap environment into itself, which greatly simplifies
further development. At the bottom of Isabelle/ML is
Poly/ML\footnote{\url{https://www.polyml.org}} by David Matthews. Started in 1985,
Poly/ML and has gone through many phases of improvements and further
scaling, often specifically for Isabelle. These are the main aspects of
Isabelle/ML technology (notably for scaling):

\begin{itemize}%
\item fast run-time compilation to produce fast machine-code

\item shared-memory parallelism (either low-level threads + locks, or
high-level future values)

\item stop-the-world garbage collection with internal parallelism

\item implicit substructure-sharing of pure values (e.g. relevant for strings,
terms, types)

\item dumped-world images for fast reloading of semantic state

\item compact representation of data on 64bit hardware: 32bit addressing of max.
16\,GB heap space%
\end{itemize}

After decades of performance tuning, it has become hard to imagine a
different ML platform to carry the weight of Isabelle applications: it would
mean a loss of one or two orders of magnitude in performance.%
\end{isamarkuptext}\isamarkuptrue%
\isadelimdocument
\endisadelimdocument
\isatagdocument
\isamarkupparagraph{Isabelle/Scala%
}
\isamarkuptrue%
\endisatagdocument
{\isafolddocument}%
\isadelimdocument
\endisadelimdocument
\begin{isamarkuptext}%
is a library for Isabelle system programming based on regular Scala, which
is hosted on the Java platform (version 11). Isabelle/Scala continues the
functional programming style of Isabelle/ML. There is an overlap of many
fundamental modules with Isabelle/ML, e.g. to manage files and processes.
The main purpose of Isabelle/Scala is to connect to the outside world in
ways beyond Isabelle/ML, e.g. TCP servers, databases, GUIs. These are the
main aspects of Isabelle/Scala technology (notably for scaling and
connectivity):

\begin{itemize}%
\item multi-threaded JVM with parallel garbage collection (but: problems with
very large heaps)

\item efficient functional programming on the JVM (but: very slow Scala
compiler)

\item access to external databases (notably SQLite, PostgreSQL)

\item access to TCP services (notably SSH, HTTP)

\item support for Mercurial (the standard SCM for Isabelle + AFP)%
\end{itemize}

Old versions of Isabelle (and other proof assistants) often had a tendency
to surround the core ML program by funny ``scripts'' for system programming
(e.g. in \isatt{bash}, \isatt{perl}, \isatt{python}, \isatt{ruby}). Isabelle/Scala avoids that:
a typed functional-object-oriented language is used instead. For example, to
explore source dependencies of Isabelle sessions (articles in AFP), there is
an underlying data structure for acyclic graphs, and explicit access to the
theory syntax in each node. Consequently, the Isabelle build tool is
implemented as a Scala module that manages source dependencies and ML
processes, without the odd \isatt{Makefile}s seen in the past.

Isabelle2019 includes 1.6\,MB of Isabelle/Scala sources; the classic ML
code-base of Isabelle/Pure has 2.4\,MB (the Scala programming style is
almost as compact as that of ML). Isabelle/Scala includes basic libraries
and concrete applications like Isabelle/jEdit (GUI application based on
AWT/Swing) or Isabelle/VSCode (Language Server Protocol based on JSON). Many
command-line tools are implemented in Isabelle/Scala as well, using an
imitation of \isatt{getopts} from GNU bash.%
\end{isamarkuptext}\isamarkuptrue%
\isadelimdocument
\endisadelimdocument
\isatagdocument
\isamarkupparagraph{Isabelle/PIDE%
}
\isamarkuptrue%
\endisatagdocument
{\isafolddocument}%
\isadelimdocument
\endisadelimdocument
\begin{isamarkuptext}%
is the Prover IDE framework of Isabelle. It is mainly implemented in
Isabelle/Scala, but also has corresponding parts in Isabelle/ML. The idea is
to treat the prover as \emph{formal document processor}, based on edits from the
front-end and markup reports from the back-end.

Isabelle/PIDE is best-known for its Isabelle/jEdit application \cite{isabelle-jedit}, which is the default Isabelle user-interface. After
download of the main desktop
application\footnote{\url{https://isabelle.in.tum.de/website-Isabelle2019}}, users are
first exposed to
Isabelle/jEdit\footnote{\url{https://isabelle.in.tum.de/website-Isabelle2019/img/isabelle_jedit.png}},
where all other tools integrated.

Isabelle2019 also provides a \emph{Headless PIDE} session: either as an
interactive object under program control in Isabelle/Scala, or as a TCP
server that understands JSON messages. That opens possibilities to digest
Isabelle libraries, based on semantic PIDE markup produced by the prover.
For example:

\begin{itemize}%
\item Export of formal content with access to the internal ML context, the
original sources, and PIDE markup over the sources (e.g. to determine
where logical constants appear in the document).

\item Update of theory sources based on PIDE markup (e.g. to replace term
notation).

\item Detailed recording of timing information over the text.%
\end{itemize}%
\end{isamarkuptext}\isamarkuptrue%
\isadelimdocument
\endisadelimdocument
\isatagdocument
\isamarkupsection{Application: Isabelle/MMT --- OMDoc and RDF/XML from AFP%
}
\isamarkuptrue%
\endisatagdocument
{\isafolddocument}%
\isadelimdocument
\endisadelimdocument
\begin{isamarkuptext}%
MMT\footnote{\url{https://uniformal.github.io}} is a language, system and library (in
Scala) to represent a broad range of languages in the OMDoc format: this
supports formal, informal, semi-formal content. The MMT repository includes
general APIs to operate on OMDoc theories, together with various tools and
applications. There are several MMT sub-projects to connect to other
systems: this includes Isabelle/MMT, implemented by myself in 2018/2019. See
also \cite[\S3.1]{CICM-2019:RDF} for a brief description of the Isabelle
theory content covered by Isabelle/MMT. See also the download
\url{https://isabelle.sketis.net/Isabelle_MMT_CICM2019}.

From the perspective of Isabelle, MMT is another system component with its
own \isatt{mmt.jar} to add modules to the Scala/JVM package namespace. This
includes the following command-line entry points:

\begin{itemize}%
\item \isatt{isabelle\ mmt{\char`\_}build} to build the MMT project inside the Isabelle
system environment;

\item \isatt{isabelle\ mmt{\char`\_}import} to import the content of a headless Isabelle/PIDE
session into MMT (OMDoc and RDF/XML triples);

\item \isatt{isabelle\ mmt{\char`\_}server} to present imported content using the built-in
HTTP server of MMT;

\item \isatt{isabelle\ mmt} to run the interactive MMT shell inside the Isabelle
system environment, e.g. for experimentation within the Isabelle + MMT
package namespace, using the \isatt{scala} sub-shell.%
\end{itemize}

The main functionality is provided by \isatt{isabelle\ mmt{\char`\_}import}: that is a
medium-sized Scala module (57KB) within the MMT code-base. It refers to
general export facilities of Isabelle/Scala, which are part of the
Isabelle2019 release. The overall tool setup is as follows:

\begin{itemize}%
\item Command-line arguments similar to \isatt{isabelle\ build} allow to specify a
sub-graph of Isabelle sessions (e.g. everything from AFP, excluding the
\isatt{very{\char`\_}slow} group).

\item A headless PIDE session in Isabelle/Scala is created, with Isabelle/Pure
as logical basis. All theories from the specified sessions are given to it
as one big document edit.

\item The PIDE session continuously processes the overall theory graph;
whenever a node is finished (including all its imported nodes), the result
is ``committed'' by a Scala operation that traverses its semantic content
and produces OMDoc and RDF/XML accordingly. Committed nodes are removed
from the PIDE session eventually, to free resources of the running ML +
Scala process.%
\end{itemize}

The headless PIDE session allows interaction under program control. Compared
to classic batch-mode (e.g. \isatt{isabelle\ build}), it requires more resources
within a single ML process, managed by a single Scala process. To digest
Isabelle + AFP including \isatt{slow} / \isatt{large} excluding \isatt{very{\char`\_}slow} sessions,
the two processes require approx. 30\,GB memory each. Note that the degree
of parallelism is diminished: the hardware configuration from
\secref{sec:motivation} with 8 processes / 8 threads is able to crunch
everything in less than 2h, but a single PIDE session requires a full day.
In return, there is one big document for all of Isabelle + AFP, with full
access to semantic states and PIDE markup.

Isabelle/MMT is the first non-trivial application of headless PIDE, and
there is further potential for performance tuning, e.g. by exploiting the
incremental nature of PIDE processing (based on edits). Presently, the
single document edit for thousands of theories can keep the prover busy for
approx. 1h, just to digest the outline of theory and proof commands.

In the future we shall see further improvements, and eventually a
convergence of batch-builds and PIDE processing, such that \isatt{isabelle\ build}
may work with semantic document markup routinely (e.g. for advanced HTML
presentation).%
\end{isamarkuptext}\isamarkuptrue%
\isadelimtheory
\endisadelimtheory
\isatagtheory
\isacommand{end}\isamarkupfalse%
\endisatagtheory
{\isafoldtheory}%
\isadelimtheory
\endisadelimtheory
\end{isabellebody}%

\bibliographystyle{abbrv}
\bibliography{root}

\begin{thebibliography}{1}

\bibitem{CICM-2019:RDF}
A.~Condoluci, M.~Kohlhase, D.~M\"uller, F.~Rabe, C.~Sacerdoti~Coen, and
  M.~Wenzel.
\newblock Relational data across mathematical libraries.
\newblock In C.~Kaliszyk, E.~Brady, A.~Kohlhase, and C.~Sacerdoti~Coen,
  editors, {\em Intelligent Computer Mathematics (CICM 2019)}, volume 11617 of
  {\em Lecture Notes in Artificial Intelligence}. Springer, 2019.

\bibitem{paulson700}
L.~C. Paulson.
\newblock {Isabelle}: The next 700 theorem provers.
\newblock In P.~Odifreddi, editor, {\em Logic and Computer Science}, pages
  361--386. Academic Press, 1990.

\bibitem{isabelle-jedit}
M.~Wenzel.
\newblock {\em {Isabelle/jEdit}}, 2019.
\newblock Documentation for Isabelle2019 (June 2019),
  \url{https://isabelle.in.tum.de/website-Isabelle2019/dist/Isabelle2019/doc/jedit.pdf}.

\end{thebibliography}

\end{document}